\documentclass[conference]{IEEEtran}

\IEEEoverridecommandlockouts

\usepackage{cite}
\usepackage{amsmath,amssymb,amsfonts}
\usepackage{algorithmic}
\usepackage{algorithm}
\usepackage{graphicx}
\usepackage{textcomp}
\usepackage[table]{xcolor}
\usepackage{tikz}
\usetikzlibrary{shapes,arrows,positioning,fit,calc,arrows.meta}
\usepackage{booktabs}
\usepackage{multirow}
\usepackage{listings}
\usepackage{microtype}
\usepackage{hyperref}
\usepackage[capitalise]{cleveref}

\hypersetup{pdfauthor={},pdfsubject={},pdfkeywords={},
            pdftitle={A Rowhammer attack on Fil-C's software capabilities}}

\colorlet{attcol}  {red!13}
\colorlet{viccol}  {blue!10}
\colorlet{hwcol}   {gray!12}
\colorlet{rowAcol} {red!20}
\colorlet{rowVcol} {green!17}
\colorlet{slotempty}{gray!10}
\colorlet{slotfull} {blue!20}
\colorlet{slottgt}  {red!18}

\tikzset{
pbox/.style={rectangle, draw=black!50, rounded corners=2pt,
               minimum height=0.70cm, text centered, text width=1.72cm,
               font=\scriptsize},
  attbox/.style={pbox, fill=attcol,  draw=red!40},
  vicbox/.style={pbox, fill=viccol,  draw=blue!35},
  hwbox/.style ={pbox, fill=hwcol,   draw=gray!45},
arr/.style ={->, >=stealth, thick},
  darr/.style={->, >=stealth, thick, dashed},
slt/.style={draw=black!45, minimum width=0.52cm, minimum height=0.40cm,
              inner sep=0, font=\tiny},
drow/.style={draw=black!50, minimum height=0.62cm, font=\tiny,
               text centered}
}

\newcommand{\FilC}{Fil-C}
\newcommand{\InvisiCap}{InvisiCap}
\newcommand{\InvisiCaps}{InvisiCaps}

\newcommand{\code}[1]{\texttt{#1}}
\newcommand{\zgetlower}{\code{zgetlower}}
\newcommand{\zgetupper}{\code{zgetupper}}
\newcommand{\zlength}{\code{zlength}}

\newcommand{\payload}{\code{payload}}

\newcommand{\payloadhdr}{\payload\ensuremath{-16}}

\newcommand{\zerotoone}{\ensuremath{0\!\rightarrow\!1}}
\newcommand{\expand}{\ensuremath{2^k}}

\usepackage{ifthen}
\newboolean{showcomments}
\setboolean{showcomments}{false}
\ifthenelse{\boolean{showcomments}}
  {\newcommand{\mynote}[2]{
    \fbox{\bfseries\sffamily\scriptsize#1}
    {\small$\blacktriangleright$\textsf{\emph{#2}}$\blacktriangleleft$}
   }
  }
  {\newcommand{\mynote}[2]{}
  }

\newcommand{\CF}{cf.\ }
 
\begin{document}

\title{A Rowhammer attack on Fil-C's software capabilities}

\author{
  \IEEEauthorblockN{Amir Naseredini}
   \IEEEauthorblockA{Independent Researcher, Cambridge, UK
    \\ sahnaseredini@gmail.com }

  \and
  \IEEEauthorblockN{Suhas Viswanath}
  \IEEEauthorblockA{Independent Researcher, Cambridge, UK\\ suhaskv0111@gmail.com}

  \and
  \IEEEauthorblockN{Martin Berger}
  \IEEEauthorblockA{ \textit{University of Sussex \& Montanarius Ltd}
    \\ Brighton \& London, UK
    \\ contact@martinfriedrichberger.net }
}
 \maketitle

\begin{abstract}

  \FilC{} retrofits spatial and temporal memory safety onto C/C++
  using software capabilities (bounds that are checked at
  every pointer dereference). \FilC{}'s guarantee rests on an unstated
  hardware assumption: that DRAM preserves the capabilities.  We show
  that Rowhammer breaks it. In \FilC{}~0.679 a part of each
  heap pointer's capability lives in a 16-byte header stored inline in
  ordinary DRAM immediately before the object pointed to, with no
  mechanism to prevent capabilities from corruption. A single \zerotoone{} bit flip
  in the capability's upper-bound expands the pointer's permitted
  range by \expand{}\,bytes and is invisible to the runtime, raising
  no panic or signal. We demonstrate this by using \FilC{} to
  \emph{convert} existing Rowhammer attacks into concrete attacks
  against a heterogeneous collection of victim software. To aid
  reproducibility (not everybody has
  DRAM vulnerable to Rowhammer) we split our demonstration
  into two paths. First, along a \emph{deterministic path},
  we isolate \FilC{} as a novel attack vector
  using a deterministic, Rowhammer-like fault-injection model. Across
  four case studies, this model corrupts the upper-bound word of a
  pointer, permitting previously out-of-bounds reads and writes.  We
  use this to achieve a bounded read/write window, secret disclosure,
  privilege escalation by flipping an adjacent \texttt{is\_admin}
  field, and the leakage of a tenant API key in a multi-tenant
  credential service backed by SQLite (compiled under
  \FilC{}). Second, along a \emph{physical end-to-end path}, we
  demonstrate all four attacks on real hardware equipped
  with DDR3 memory, and with Target Row Refresh (TRR) equipped
  DDR4, proving the exploits are physically realisable.
  The latter attack shows that the TRR defence does not close the
  \FilC{}~delivery channel.  We close with a discussion of
  \FilC-specific mitigations.

\end{abstract}
 
\section{Introduction}

Memory-safety violations in C/C++ remain the dominant root cause of
critical security vulnerabilities. Because rewriting the vast deployed base of
C/C++ software is impractical, a major line of defence retrofits spatial and
temporal memory-safety protections onto existing code. These include compiler
instrumentation as in SoftBound/CETS~\cite{softbound} and
AddressSanitizer~\cite{asan}, hardware-assisted capabilities as in
CHERI~\cite{woodruff2014}, and memory tagging as in Arm Memory Tagging Extension
(MTE)~\cite{mte}. \FilC{}~\cite{filc} is a recent and unusually complete point in
this space, instrumenting every pointer operation at compile time and enforcing
\emph{complete} spatial and temporal bounds checking at runtime through software
capabilities, with enough C/C++ compatibility to build large real-world programs
unmodified. A tacit assumption underlies such systems, that the hardware
faithfully preserves the metadata on which every check depends.

The Rowhammer vulnerability~\cite{kim2014, mutlu2023} invalidates exactly this
assumption for commodity DRAM. By repeatedly activating adjacent memory rows,
an attacker induces controlled bit flips in victim rows without ever directly
accessing them. Rowhammer has corrupted OS page
tables~\cite{seaborn2015}, cryptographic keys~\cite{ffs}, deduplication and
allocator state of co-located VMs~\cite{ffs,drammer}, and JavaScript
sandboxes~\cite{gruss2016, smash2021}. Rowhammer reads secrets it
never touches~\cite{rambleed} and survives across DRAM generations and
defences~\cite{blacksmith, rowpress, trrespass, zenhammer}. Yet
compiler-managed bounds metadata, the very state on which software-defined
memory safety rests, has, to our knowledge, received no systematic analysis as
a Rowhammer target. This paper fills that gap with an explicitly
\emph{conditional} thesis:
\begin{quote}
  \emph{Wherever Rowhammer can land a usable bit flip in a suitable
  part of the stored metadata associated with a pointer, \FilC{} cannot
  hold its spatial-safety promise for that pointer}.
\end{quote}
We are \emph{not} proposing a new way to hammer DRAM. We show what an existing,
widely demonstrated hardware fault means for a software defence that relies on
the tacit assumption above. We exploit a design choice in \FilC{}~0.679, where
part of each heap pointer's capability is stored as a 16-byte header inline in
ordinary DRAM immediately before the object, with no mechanism protecting it from
corruption. Rowhammer does not let an attacker pick an arbitrary bit and force it
to flip. It identifies physically vulnerable cells and grooms valuable victim data
onto one of them~\cite{seaborn2015,ffs}, and most flips are unusable, ruled out by
the cell's page offset, its position within the metadata, the flip polarity, the
bit's current value, or the requirement that the result remain a valid pointer
(\S\ref{sec:phys_feas}). Our attack needs only \emph{one} flip that survives all
these filters, and such a flip silently defeats the \FilC{} runtime with no panic,
signal, or machine check to mark it.

Our attack pipeline has three components: Rowhammer fault generation, grooming the
victim to align \FilC{}'s unprotected capabilities with vulnerable cells, and
silently circumventing \FilC{}'s runtime checks once a bit flip expands the
capability's upper bound.

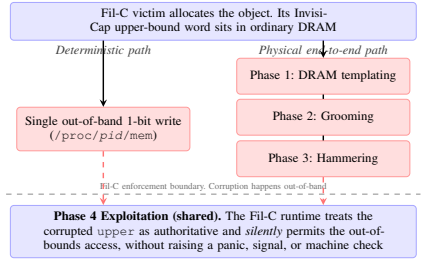
\begin{figure}[t]
\centering
\resizebox{0.62\columnwidth}{!}{\begin{tikzpicture}[font=\scriptsize, >=stealth,
   lane/.style={font=\scriptsize\itshape, text=black!75},
   note/.style={font=\tiny, text=black!55}]

\node[vicbox, text width=7.7cm, minimum height=0.60cm] (start) at (0,4.55)
       {\FilC{} victim allocates the object. Its \InvisiCap{} upper-bound word
      sits in ordinary DRAM};

\node[lane] at (-2.15,3.98) {Deterministic path};
\node[lane] at ( 2.15,3.98) {Physical end-to-end path};
  \node[note] at ( 2.15,3.50) {proves delivery};

\node[attbox, text width=3.15cm, minimum height=0.85cm] (det) at (-2.15,2.45)
     {Single out-of-band 1-bit write\\(\texttt{/proc/\textit{pid}/mem})};

\node[attbox, text width=3.15cm] (p1) at (2.15,3.48) {Phase 1: DRAM templating};
  \node[attbox, text width=3.15cm] (p2) at (2.15,2.68) {Phase 2: Grooming};
  \node[attbox, text width=3.15cm] (p3) at (2.15,1.87) {Phase 3: Hammering};
  \draw[arr,-](p1)--(p2); \draw[arr,-](p2)--(p3);

\draw[dashed,black!45] (-4.05,1.18) -- (4.05,1.18);
  \node[note] at (0,1.33) {\FilC{} enforcement boundary. Corruption happens out-of-band};

\node[vicbox, text width=7.7cm, minimum height=0.95cm] (cons) at (0,0.45)
     {\textbf{Phase 4 Exploitation (shared).}  The \FilC{} runtime treats the
      corrupted \texttt{upper} as authoritative and \emph{silently} permits the
      out-of-bounds access, without raising a panic, signal, or machine check};

\draw[arr](start.south-|det.north) -- (det.north);
  \draw[arr](start.south-|p1.north)  -- (p1.north);
\draw[darr,red!55](det.south) -- (det.south|-cons.north);
  \draw[darr,red!55](p3.south)  -- (p3.south|-cons.north);

\end{tikzpicture}}
\caption{The two experimental paths.  Blue boxes run inside the \FilC{} victim
process. Red boxes run in the attacker's native (non-\FilC{}) process.  Both
paths share the same start (a live \InvisiCap{} in DRAM) and the same
consequence (Phase~4: the runtime silently honours the corrupted bound). They
differ only in \emph{delivery}.  The \emph{deterministic path} lands the flip
with a single out-of-band write to isolate the metadata-corruption consequence.
The \emph{physical end-to-end path} lands it by real Rowhammer through
templating, grooming, and hammering (Phases~1--3).  Neither delivery is
mediated by a \FilC{} check, which is why the corruption is invisible to the
runtime.}
\label{fig:pipeline}
\end{figure}
 
\noindent To aid reproducibility and cleanly separate these components, our
paper proceeds along two complementary paths (Figure~\ref{fig:pipeline}).
First, a \emph{deterministic path} isolates the circumvention
stage using a deterministic, Rowhammer-like fault-injection
model. This allows us to study the impact of targeted bit flips on
\FilC{} binaries independently of the underlying hardware, yielding
four concrete case studies: a bounded out-of-bounds (OOB) read/write window,
secret disclosure, privilege escalation by flipping an adjacent
\texttt{is\_admin} field, and cross-tenant API key leakage in a
multi-tenant credential service backed by SQLite. Second, a
\emph{physical end-to-end path} proves that the necessary fault
generation and grooming are physically realisable by executing these
attacks end-to-end on commodity hardware.
Our evaluation uses both older DDR3 memory (exploited in
2 minutes) and modern, Target Row Refresh (TRR) protected DDR4 modules
(exploited in 10--35 minutes), demonstrating that 
the TRR defence does not close  this \FilC{} delivery channel.

We emphasise that mitigating Rowhammer is not \FilC{}'s responsibility: it is a
hardware-level defect for DRAM vendors and platform designers. However, Rowhammer
is widely and incorrectly believed to be solved, so it is worth stating plainly
that a software safety system such as \FilC{} remains exposed to it invisibly to
the runtime. We close with \FilC{}-specific mitigations: we contribute no new
class of defence, and the Message Authentication Code (MAC) protection we
recommend is PT-Guard's~\cite{saxena2023} transported to a new target.

In summary, our contributions are as follows.
\begin{itemize}

  \item We present the first Rowhammer attack on \FilC{}.  We
    characterise the \FilC{}~0.679 capability layout and show that its
    bounds checks rely on the integrity of inline metadata stored in
    ordinary DRAM (\S\ref{sec:layout}, \S\ref{sec:assumptions},
    \S\ref{sec:related}).

  \item Through deterministic single-bit fault injection, we show that
    corrupting software capabilities silently bypasses \FilC{}'s
    runtime checks.  Four case studies demonstrate consequences including
    privilege escalation and cross-tenant information disclosure
    (\S\ref{sec:poc}, \S\ref{sec:eval}).

  \item We develop a capability-aware Rowhammer delivery method combining
    flip selection with allocator- and offset-aware physical-frame
    grooming.  We use it to realise the attacks end-to-end on DDR3 and
    the tested TRR-equipped DDR4 system, demonstrating that TRR did not
    prevent delivery on that platform (\S\ref{sec:attack},
    \S\ref{sec:phys_feas}, \S\ref{sec:ddr4_phys}).

  \item We compare \FilC{}'s exposure with hardware-tagged capability
    systems such as CHERI and propose \FilC{}-specific mitigations
    (\S\ref{sec:related}, \S\ref{sec:mit}).

\end{itemize}
 \section{Background}\label{section_background}

This section reviews the relevant background about \FilC{} with its
software capability system and non-moving garbage collector, as well
as key facts about Rowhammer.

\subsection{\FilC{}, \InvisiCaps{} and FUGC}
\label{sec:layout}

\usetikzlibrary{decorations.pathreplacing,positioning}

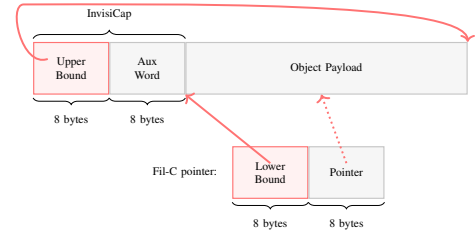
\begin{figure}[t]
\centering
\resizebox{0.70\columnwidth}{!}{\begin{tikzpicture}[
    font=\scriptsize,
    box/.style={
        draw=gray!45,
        fill=gray!7,
        line width=0.7pt,
        minimum height=1.1cm,
        align=center
    },
    ptrbox/.style={
        draw=red!55,
        fill=red!5,
        line width=0.7pt,
        minimum height=1.1cm,
        align=center
    },
    pointer/.style={
        ->,
        very thick,
        red!55
    }
]

\node[ptrbox, minimum width=1.6cm] (upper) {Upper\\Bound};
\node[box, minimum width=1.6cm, right=0pt of upper] (aux) {Aux\\Word};
\node[box, minimum width=6cm, right=0pt of aux] (payload) {Object Payload};

\node[ptrbox, minimum width=1.6cm, below=1.1cm of payload.south west, xshift=1cm, anchor=north west] (lower) {Lower\\Bound};
\node[box, minimum width=1.6cm, right=0pt of lower] (intval) {Pointer};

\foreach \block in {upper,aux,lower,intval} {
    \draw[
        decorate,
        decoration={brace,mirror,amplitude=4pt}
    ]
        ([yshift=-2pt]\block.south west) --
        ([yshift=-2pt]\block.south east)
        node[midway, below=7pt] {8 bytes};
}

\draw[
    decorate,
    decoration={brace,amplitude=5pt}
]
    ([yshift=2pt]upper.north west) --
    ([yshift=2pt]aux.north east)
    node[midway, above=8pt] {\InvisiCap{}};

\node[left=0.2cm of lower, font=\scriptsize] {Fil-C pointer:};

\draw[pointer] ([yshift=0.2cm]lower.center) -- (payload.south west);

\draw[dotted,pointer] ([yshift=0.2cm]intval.center) -- ([xshift=2.9cm]payload.south west);

\coordinate (upper-route) at
    ([xshift=-0.2cm,yshift=0.8cm]upper.north west);
\draw[pointer]
    ([xshift=-0.5cm,yshift=0.2cm]upper.center)
    .. controls
        ([xshift=-0.25cm]upper.west) and
        ([xshift=-0.35cm]upper-route)
    .. (upper-route)
    .. controls
        ([xshift=2.9cm]upper-route) and
        ([yshift=0.9cm]payload.north east)
    .. (payload.north east);

\end{tikzpicture}}

\caption{Visualisation of a \FilC{} pointer and capability layout.}\label{fig:invisiCaps}

\end{figure}
 
\FilC{} is a memory-safe implementation of C and C++ that remains compatible with
advanced idioms such as multithreading and atomics, and whose design draws on
SoftBound~\cite{softbound} and CHERI~\cite{CHERI_specv0.9.8.2}. It associates every
pointer with a software capability, an \InvisiCap{}, recording the object's upper
bound, and on each dereference the runtime checks the access against this bound and
the lower bound carried by the pointer itself (Figure~\ref{fig:invisiCaps}). The
capability is \emph{invisible} in that programs can neither create nor forge it:
the compiler and runtime maintain it, yet it never appears in the pointer's C-level
value. Physically it is nothing exotic. For a heap object it is a plain 16-byte
header stored inline in ordinary DRAM, and this gap between logical invisibility
and physical ordinariness is exactly what we exploit. We fix three terms used
throughout the paper because the attack acts on each differently. Following the
C/C++ standard we call the whole allocation the \emph{object}. Within it,
\emph{capability} is the metadata at a negative offset from the concrete pointer
(the inline header recording the bounds) and
\emph{payload} is the program-visible data at a non-negative offset. Concretely,
the 16-byte header holds the 8-byte upper bound (\texttt{upper}) at \payloadhdr{},
which is our target, and an 8-byte aux word at \texttt{payload}$-8$ that the
attack does not use, with the object payload beginning at \texttt{payload}$+0$.

\noindent
We empirically confirm this layout in \FilC{}~0.679 by allocating objects with
the toolchain and querying the runtime's capability introspection API
(\zgetlower, \zgetupper, and \zlength)~\cite{PizloF:capintapi}. Corrupting the
word at \payloadhdr{} out of band makes \zgetupper\ return the corrupted value
and \zlength\ report the correspondingly enlarged size (\S\ref{sec:eval}).
Crucially, this word is plain DRAM, with no MAC, parity, or separate tag memory
protecting it. \FilC's runtime check before a dereference is essentially the
following:
\begin{lstlisting}
if (access_addr < cap.lower || access_addr >= cap.upper)
    filc_panic("out-of-bounds");
\end{lstlisting}
A corrupted value normally panics with ``\texttt{cannot read/write pointer with
ptr >= upper}'', and the check fails to protect the object only if \texttt{upper}
has been silently modified outside the runtime's view.

\FilC's garbage collector, FUGC (``Fil's Unbelievable Garbage
Collector''~\cite{PizloF:filunbgc}), is another component our attack relies on.
It is an accurate, concurrent, \emph{non-moving} mark-and-sweep collector. For
small objects its \texttt{libpas}-based user-space allocator groups allocations
by size class into slab-style allocator pages and returns reclaimed slots to
size-class-local free lists~\cite{Bonwick1994SlabAllocator,LeijenEtAl2019Mimalloc},
which should not be confused with Linux's kernel slab allocator. FUGC determines
the stable virtual layout of objects within an allocator page, whereas Linux
determines which physical frame backs that virtual page, so our physical
placement relies on Linux-specific behaviour, including page-frame recycling,
memory-zone locality, and the \texttt{/proc/\textit{pid}/pagemap} interface used
to observe placement (\S\ref{sec:poc_phys}). The deterministic path
(\S\ref{sec:attack}) additionally uses \texttt{/proc/\textit{pid}/mem}. Two properties of FUGC matter here. First, because FUGC is non-moving, a
capability remains at a fixed \emph{virtual} address while its object is live,
removing collector-induced relocation from the attack, though Linux may still
migrate the underlying physical frame, so the physical harness must revalidate that
mapping. Second, neither allocation nor the sweep re-verifies the bounds against a
stored checksum, so a corrupted bound persists across collection cycles. Phase~2
(\S\ref{sec:phase2}) exploits the predictable placement, and the absence of
re-verification lets the injected corruption survive.

\subsection{DRAM and Rowhammer}

Each DRAM cell holds a bit as charge in a capacitor that
leaks over time, and must be refreshed regularly.
Rowhammer~\cite{kim2014} exploits the electromagnetic
coupling between physically adjacent rows: repeatedly activating an
\emph{aggressor} row accelerates charge leakage in its neighbours, and
once enough activations accumulate before the next refresh, a bit in a
\emph{victim} row flips.  Double-sided hammering places aggressors on
both sides of the victim, doubling the disturbance rate and
inducing flips within one refresh interval.
A decade of subsequent work has grown this circuit-level effect into a family of
attacks and defences, catalogued in recent surveys~\cite{rhsurvey,mutlu2019retro}.
We use one distinction from that literature, between the physical \emph{delivery}
of a flip and the \emph{consequence} an attacker constructs from it, a separation
the survey~\cite{rhsurvey} makes explicit. Our contribution lies on the
consequence side. For delivery we rely on established patterns, so the result is
not tied to any single hammering method.
The two generations we target differ in their mitigations. DDR3 has none beyond
standard refresh, whereas DDR4 adds in-DRAM TRR, which samples row activations and
pre-emptively refreshes neighbours. We bypass them with classic double-sided and
many-sided hammering respectively (\S\ref{sec:attack}). Both use an 8\,KiB row
size, and off-die ECC is not a DRAM-intrinsic mitigation.
 \section{Threat model}
\label{sec:threat}

\subsection{Attacker capabilities}

The attacker runs a process co-resident with a \FilC{} victim, aiming to induce a
single bit flip in the DRAM cell holding the victim's \InvisiCap{} upper bound. The
key ability it needs is to modify that cell \emph{out-of-band}, from outside any
\FilC{}-instrumented instruction, so that no \FilC-inserted check detects the
change. Physical Rowhammer provides exactly this and has been demonstrated from
unprivileged userspace~\cite{seaborn2015}, from JavaScript without
\texttt{clflush}~\cite{gruss2016,smash2021}, and across DRAM
generations~\cite{kim2014,blacksmith,rowpress}, so we do not treat privilege as the
crux of our contribution. We address this along two paths, deterministic then
physical end-to-end (\CF Figure~\ref{fig:pipeline}).

\subsubsection*{Deterministic path (out-of-band injection)}
Our deterministic fault-injection experiment (\S\ref{sec:poc},
\S\ref{sec:eval}) runs the victim with the \FilC{} 0.679 toolchain and, from a
same-UID parent authorised to trace its own dumpable child
(\texttt{kernel.yama.ptrace$\_$scope\, =\, 1}), writes a value differing from
the original bound in exactly one bit through \texttt{/proc/\textit{pid}/mem}.
This is not an end-to-end Rowhammer attack, because the parent already holds
arbitrary access to the child's address space. It is instead a reproducible
experiment showing that, once the one-bit corruption is visible to the victim,
the complete \FilC{} runtime accepts an access the original bound would reject.
Physical delivery is evaluated separately in \S\ref{sec:poc_phys}.

\subsubsection*{Physical end-to-end path (bare-metal hammering)}
We provide a bare-metal hammering engine (\S\ref{sec:poc_phys}) that targets
the victim's \InvisiCap{} word by its physical address, recovered from
\texttt{/proc/\textit{pid}/pagemap}~\cite{linux_pagemap}, and hammers its
attacker-owned neighbours. We run it with classical double-sided hammering on
our DDR3 lab machine and, with wide many-sided TRR-evading hammering and
physical-frame grooming, end-to-end on a second, TRR-equipped DDR4 lab machine
(\S\ref{sec:eval}, \S\ref{sec:ddr4_phys}). That a usable flip is achievable on
susceptible DDR3 and DDR4 we confirm ourselves (\S\ref{sec:phys_feas},
\S\ref{sec:ddr4_phys}), and the broader Rowhammer literature establishes it more
generally~\cite{kim2014,blacksmith}.

\subsection{Attack preconditions}
\label{sec:assumptions}
\FilC's stated software memory-safety guarantee assumes the hardware faithfully
stores the capability metadata maintained by trusted compiler and runtime code.
We deliberately relax that hardware-integrity assumption and ask not whether some
\FilC{} instruction is unchecked, but whether \FilC{} detects corruption
introduced below its software enforcement.  That assumption unpacks into several
concrete properties, none stated by \FilC{}, because a software runtime checks
the invariants of its own code and trusts the storage beneath it.  Each below is
a property the guarantee needs but never verifies, annotated with whether
Rowhammer upholds or breaks it.

\begin{itemize}

  \item \textbf{Metadata immutability.}
\InvisiCap{} fields change only via the \FilC{} runtime.
\emph{Violated} by Rowhammer flips.

\item \textbf{Allocator opacity.}
The FUGC slab layout is unpredictable to outside attackers.
\emph{False}: the slab is deterministic given fixed heap state, which Phase~2
(\S\ref{sec:phase2}) exploits directly.

\item \textbf{Physical isolation.}
Attacker-controlled DRAM rows are never adjacent to victim \InvisiCap{} pages.
\emph{False}: Linux provides no such guarantee.

\item \textbf{Refresh regularity.}
Each row is refreshed often enough to stop disturbance from accumulating.
\emph{Violated}: Rowhammer accumulates enough activations within a single 64\,ms
window to flip a bit, and both machines run at their standard refresh rate.

\item \textbf{ECC coverage.}
The victim's DRAM applies no error correction, so a single-bit flip is neither
detected nor corrected. Both machines use non-ECC DIMMs. SECDED ECC would catch
the lone flip, and defeating it requires crafted \emph{multi-bit} patterns with
additional templating and fault handling~\cite{eccploit,didio2023}, which we
leave as future work.

\end{itemize}
 \section{Attack methodology}
\label{sec:attack}

We now detail the two complementary experimental paths, deterministic and
physical end-to-end, from Figure~\ref{fig:pipeline}. They share the same start, a
live \InvisiCap{} in the \FilC{} victim's DRAM, and the same consequence, Phase~4,
in which the runtime silently honours the corrupted bound, and diverge only in
\emph{delivery}. The physical end-to-end path runs as a four-phase pipeline: an
attacker process compiled with standard GCC executes Phases~1--3 against physical
DRAM, and the corrupted \InvisiCap{} then causes the \FilC{} victim to silently
permit out-of-bounds access in Phase~4. The deterministic path collapses
Phases~1--3 into a single out-of-band write and shares Phase~4 unchanged. Beyond
identifying compiler-managed bounds metadata as a target, we contribute a
capability-aware delivery methodology that combines established primitives (Flip
Feng Shui frame donation~\cite{ffs}, DRAMA bank recovery~\cite{pessl2016drama},
and many-sided hammering~\cite{trrespass,blacksmith}) into a workflow for landing
a \emph{usable} flip in a capability word, adding memory-zone-aware scouting,
offset-aware frame selection, and a per-candidate no-hammer control that separates
hammer-induced flips from unrelated memory errors.

\subsection{Phase 1: DRAM templating}
\label{sec:phase1}

Phase~1 locates a \emph{flippy} cell (one that reliably flips \zerotoone{}
under hammering) satisfying two conditions: (i)~the cell maps to the 8-byte upper-bound word of an
object's inline header (at \payloadhdr{} once a capability is
groomed onto it, \S\ref{sec:phase2}); (ii)~the flipped bit is at a position
$k$ large enough that the resulting expansion $2^k$ reaches the attacker's
intended target (a few hundred bytes suffices to reach an adjacent object,
larger $k$ yields a wider window).

\subsubsection*{Memory mapping and geometry recovery}

To hammer, the attacker needs a pool in which a virtual row offset maps to the same
physical row offset, or the DRAMA-recovered bank geometry does not hold. The engine
backs its hammer pool with transparent huge pages (THP, 2\,MiB) for that
contiguity, disabling them only for grooming (\S\ref{sec:phase2}), where engine and
forked victim allocate at the same 4\,KiB granularity so a single donated frame is
reused. Physical frame numbers are recovered via \texttt{/proc/self/pagemap}, and
DRAM bank XOR functions at runtime via the DRAMA timing
method~\cite{pessl2016drama}, where same-bank address pairs exhibit a $\sim$30\,ns
bank-conflict penalty. The engine recovers the geometry directly from this timing
channel rather than any hardcoded mapping, so the routine carries over unchanged
across our DDR3 and DDR4 machines.

\subsubsection*{Generation-specific hammering}

\textbf{DDR3: classic double-sided} (\CF \S\ref{sec:eval}).
The DDR3 modules we work with have no TRR, so a straightforward hammer loop
suffices. Each burst issues 2.5\,M iterations in our build, that is 5\,M row
activations across the aggressor pair. On our DDR3-1333 module a burst of this
size completes well within the 64\,ms refresh window, so disturbance accumulates
between refreshes.

\textbf{DDR4: wide many-sided hammering} (\CF \S\ref{sec:ddr4_phys}).
DDR4 adds TRR, an in-DRAM defence.   Blacksmith~\cite{blacksmith}
defeats TRR by \emph{fuzzing} non-uniform access patterns over frequency,
phase, and amplitude until one evades the (undocumented, vendor-specific)
sampler. Our preset instead uses wide many-sided hammering. It
combines the true aggressor pair with a large set of decoy rows,
distributing the sampler's limited tracking budget across many
rows. We do not claim that this preset reproduces Blacksmith's full
non-uniform search. Instead, we evaluate it end-to-end on a
TRR-equipped DDR4 module, including its effect on the runtime
(\S\ref{sec:ddr4_phys}). On this module, the preset successfully
delivers usable flips. Orthogonal read-disturbance amplifications such
as RowPress~\cite{rowpress} could reduce the activations a flip needs, but our
result does not depend on them and we do not employ them in the evaluated
preset.

\subsection{Phase 2: grooming}
\label{sec:phase2}

For the deterministic case studies the victim is an experimental harness rather
than an unmodified application: it publishes the target address and retries
allocation until the required adjacency is obtained, making the
metadata-corruption consequence reproducible. Forcing that placement without such
cooperation is the task of the physical grooming described later. FUGC hands out
same-size-class objects from contiguous slab pages at consecutive increasing
addresses drawn from the free-list head, and the physical \emph{stride} between
successive objects is what makes a target and a chosen neighbour land adjacently,
so we measured it directly.

FUGC uses the \texttt{libpas} \texttt{bmalloc} heap configuration.  A request of $r$ bytes
is rounded up to a payload-capacity class $C(r)$, and \FilC{} stores its 16-byte
\InvisiCap{} header inline immediately before the payload, so the physical stride
between successive objects is
\begin{equation}
  \mathrm{stride}(r) = C(r) + 16.
  \label{eq:stride}
\end{equation}
The header is added \emph{after} class selection, not before. We measured $C(r)$
directly rather than deriving it. On our \FilC{}~0.679 build the payload-capacity
classes are $\{\ldots, 208, 240, 288, 336, 400, \ldots\}$, a 48-byte step across
the $240$--$336$ band our objects fall in but not a single uniform step globally
(for example $336\!\to\!400$ is a 64-byte jump), and the physical stride follows
as $C(r)+16$. Independently, the runtime rounds the \emph{enforced} bound up to 16
bytes, so \zlength{} returns $240, 256, 272, 288, \ldots$ as $r$ grows, equalling
the request exactly for our 16-aligned 256-byte objects. We obtained this by
instrumenting consecutive allocations and reading back addresses and bounds, not
from the \texttt{libpas} source. The attack relies only on the measured
determinism within a size class, where a fixed stride makes adjacency predictable,
not on the global progression following any closed-form rule.

Our case-study objects are 256-byte requests, hence the 288-byte class and the
304-byte stride.  The 32-byte slack between the \InvisiCap{} upper bound (which
stops at the requested 256 bytes) and the 288-byte class capacity is inert under
an intact runtime, but it lies within the same physical slot, so even a flip that
reaches only into that slack still expands the usable window.  Because the class is fixed, the target and a chosen secondary object always
land adjacently.

\subsubsection*{Aligning the objects (deterministic case studies)}
For the reproducible case studies the alignment is performed by the victim
itself: it allocates the target and, in the cross-object cases
(\texttt{secret}/\texttt{authflag}/\texttt{kv}), a secondary object of the same
size class.  Because FUGC hands out consecutive, increasing addresses within a
slab, the secondary lands immediately above the target. The victim defensively
discards (frees and retries) any candidate that does not, which can only happen
across a slab boundary.  No slab grooming is required here, because the flip is
delivered out-of-band directly to the published upper-bound word: the target's
cell is at a fixed offset from its payload,
$v_\mathrm{cap} = v_\mathrm{payload} - 16$, so the address the victim publishes
fixes exactly the cell the attacker must flip.

\subsubsection*{Landing a physical flip (Flip Feng Shui grooming)}
When the flip is delivered by real hammering (\S\ref{sec:phys_feas},
\S\ref{sec:ddr4_phys}), owning the victim row's exact physical neighbours is
impractical, so the end-to-end path uses page-frame grooming in the Flip Feng Shui
style~\cite{ffs} (\S\ref{sec:poc_phys}).  The attacker holds the flippy \emph{cell}
fixed, templating known-flippy physical frames in its own memory and
\emph{donating} them to the allocator, then drives the victim to spray many
same-size-class objects until one \InvisiCap{} upper-bound word lands on a donated
frame.

\emph{What ``donation'' is, concretely.}
A donation is a single \texttt{munmap} of one 4\,KiB flippy frame, returning it
to the head of the running CPU's LIFO per-CPU buddy free-list while every other
attacker frame, including the physically adjacent aggressor rows, stays mapped.
The frame's physical address and its same-bank adjacency to those aggressors are
unchanged, and only its ownership moves to the allocator, from which the victim's
next same-class allocation is likely to draw it.  When a victim capability word
lands there, the attacker re-hammers the same recorded aggressors (Phase~3).

\emph{Matching the victim's page granularity.}
Grooming succeeds only if attacker and victim allocate at the same granularity.
The engine pins both to 4\,KiB base pages, disabling transparent huge pages for
itself and its forked \FilC{} victim (\texttt{prctl(PR\_SET\_THP\_DISABLE)}), so
both draw from the same order-0 buddy free-lists and a donated frame lands on the
free-list the victim's next same-class allocation pops from.  Left at its
default, THP would back both heaps with 2\,MiB pages, where a donated 4\,KiB
frame would fall to an order-0 free-list the victim's slab never touches.

\emph{Zone- and offset-aware placement.}
Because per-CPU LIFO reuse is zone-local and the victim's allocator may draw from
\texttt{ZONE\_DMA32} or \texttt{ZONE\_NORMAL} under memory pressure, the spray is
\emph{two-phase}. A small ``scout'' batch reveals the zone the victim currently
draws from, so the attacker templates and donates within that zone, and the main
spray then places capability words on the just-donated, in-zone frames.  To raise
the reuse probability without extra hammering, the attacker preferentially
donates frames whose flip falls in an exploitable low byte of the 8-byte bound
word.  A right-sized pool suffices, so reuse does not require exhausting memory,
and aggressively draining free RAM into a locked buffer is offered but disabled
by default, since it can backfire by exhausting \texttt{ZONE\_NORMAL} and pushing
the spray into \texttt{ZONE\_DMA32}.  The attacker then searches the sprayed words
for one coinciding with a flippy cell and re-hammers that frame (Phase~3).  Both
alignments exploit the same size-class determinism, differing only in whether the
object or the flippy cell is the fixed point.

\subsection{Phase 3: hammering}
\label{sec:phase3}

Phase~3 re-executes the generation-appropriate hammer pattern against the
same aggressor rows, with the \InvisiCap{} now resident in the victim row.
The hammer runs repeated bursts until the flip is observed or the time budget
expires, with an early-exit check after each burst that avoids unnecessary
activations once the flip is confirmed.

\subsubsection*{Mathematical effect}
A \zerotoone{} flip at bit $k$ of the 8-byte upper-bound word
yields:
\begin{equation}
  u' = u + 2^k
  \label{eq:expansion}
\end{equation}
For $k=30$: $u'=u+1\,\mathrm{GiB}$.  The \FilC{} check
$\mathtt{access} < u'$ then passes for every address in
$[u_\mathrm{orig},\;u_\mathrm{orig}+2^k)$, including adjacent heap
objects, function pointers, and runtime structures.

\subsection{Phase 4: exploitation}
\label{sec:phase4}

The corrupted capability is invisible to all software layers, with no panic,
machine-check exception, or signal. The \FilC{} bounds check treats the corrupted
\texttt{upper} as authoritative, so every access in $[u_\mathrm{orig},\,u')$ now
passes (\S\ref{sec:eval}). An OOB \emph{read} follows because the check
$\mathtt{base}+i<u'$ holds for $i<2^k$, letting the victim read whatever object
the allocator placed above the target, as in our adjacent-secret and SQLite
API-key case studies (the SQLite exposure is not a vulnerability in SQLite
itself, \S\ref{sec:casestudies}). The same expanded bound permits an OOB
\emph{write}, which we use to set an adjacent session object's \texttt{is\_admin}
field and escalate the victim from an unprivileged to an administrative
principal. Writing runtime-internal pointers to hijack control flow is a
plausible further escalation we do not demonstrate, but the read/write primitive
it would build on is what we show.
 \section{Proof of concept (PoC) implementations}
\label{sec:poc}

\noindent Our implementation follows the two-path model from
Fig.~\ref{fig:pipeline}. The primary PoC (\S\ref{sec:poc_real}) is what our
evaluation runs, demonstrating the consequence of Rowhammer bit flips on \FilC{}
metadata along both paths, and the bare-metal drivers (\S\ref{sec:poc_phys})
implement the physical delivery primitive the end-to-end path uses.

\subsection{Primary PoC: real-runtime bypass}
\label{sec:poc_real}

The primary PoC is a two-process pair. The \textbf{victim} (a small \FilC{}
program, or a SQLite-backed service for the \texttt{kv} case study) is compiled
with the \FilC{} 0.679 toolchain, so every access it makes is checked by the
runtime. It allocates the target object, publishes the runtime-assigned bounds
(read via \zgetlower/\zgetupper) over a small shared handshake region, waits,
then performs an access out-of-bounds for the \emph{original} object. The
\textbf{attacker}, the bare-metal engine (\S\ref{sec:poc_phys}) built with
ordinary \texttt{gcc}, lives outside the \FilC{} model. It forks the victim,
selects the case study via \texttt{--scenario} (and \texttt{--victim} for
\texttt{kv}), locates the upper-bound cell either at \payloadhdr{} or, to model
templating, by scanning the victim's writable memory for the published bound, and
applies a single \zerotoone{} flip.

The same harness supports both delivery paths: the deterministic path uses
\texttt{--selftest} to write the single-bit change directly through
\texttt{/proc/\textit{pid}/mem}, and the physical end-to-end path induces the identical
flip by hammering (\S\ref{sec:poc_phys}).
The four case studies share this harness and differ only in what the victim does
after the flip (\S\ref{sec:casestudies}). The design yields an unambiguous oracle:
if the flip took, the runtime permits the access and the victim runs to
completion, whereas in the no-flip control the first out-of-bounds access is
hard-panicked and the victim is killed before it can report. No path lets the
victim attempt an out-of-bounds access and \emph{report} that it was blocked, so a
success report can only mean the runtime permitted the access.

\subsection{Bare-metal delivery drivers}
\label{sec:poc_phys}

We implemented two generation-specific bare-metal engines, one using classical
double-sided hammering on DDR3 and the other wide many-sided hammering to evade
TRR on DDR4~\cite{blacksmith} (\S\ref{sec:attack}).  Each hammers the real
\FilC{} victim rather than a stand-in.  It forks the \FilC{}-compiled victim,
learns the \emph{physical} address of its \InvisiCap{} upper-bound word through
\texttt{/proc/\textit{pid}/pagemap}, owns that word's physical neighbours, and
hammers attacker-owned aggressor rows on both sides in a tight cache-flush loop
for a fixed wall-clock budget while polling the word for a genuine change.  The
geometry recovery, self-templating, and Flip-Feng-Shui grooming it relies on are
the Phase~1--2 primitives of \S\ref{sec:attack}, and the grooming mode produced
our physical end-to-end result (\S\ref{sec:phys_feas}).  It never simulates a
flip and reports a negative result with the measured totals.  The cache-line flush is
portable, detecting \texttt{clflushopt} via CPUID once and falling back to the
universal \texttt{clflush} (SSE2) so the loop runs on the Sandy/Ivy-Bridge DDR3
machines that are the cleanest Rowhammer targets.  Because the upper-bound word
holds a valid pointer, the engine counts a flip as \emph{exploitable} only if it
both expands the bound and preserves pointer validity, reporting any other
genuine flip as unusable.  Finally, the \texttt{--selftest} path injects the
single-bit flip through \texttt{/proc/\textit{pid}/mem} rather than a hammer,
validating the full chain on any machine, whereas the hammering campaign requires
susceptible DRAM on which the cited work establishes that flips are achievable.
 \section{Experimental evaluation}
\label{sec:eval}

\subsection{Setup}

We evaluate both paths on a commodity DDR3 lab machine (Intel Core i7-2630QM,
Sandy Bridge, 8\,GiB of DDR3 from Vendor~A, Linux 5.15), with the victim built
under \FilC{} 0.679 and the attacker with system \texttt{gcc}~11.4. The
reproducible results below (the four case studies) deliver the single-bit flip
\emph{out-of-band}, isolating the metadata-corruption consequence from physical
flip throughput. As a technique this path needs no root, no hugepages, and no
special DRAM, only the ability to ptrace a forked child, which the default
\texttt{Yama} policy permits. The \FilC{} side of every result is real and
re-verified on this machine, which also hosts the bare-metal engine
(\S\ref{sec:poc_phys}). A second, separate machine with a TRR-equipped DDR4 module
from Vendor~B is used solely for the physical result in \S\ref{sec:ddr4_phys}, and
no result mixes hardware across the two.

\subsection{Confirming the layout and the flip effect}

When we allocate a 256-byte object under the runtime, \zgetlower\ and \zgetupper\
report bounds exactly 256 bytes apart, with the upper-bound value in DRAM at
\payloadhdr{}. We apply a single \zerotoone{} flip at bit~7 of that word
($\mathtt{upper}\mathbin{+}2^7$), after which \zgetupper\ returns the corrupted
value and \zlength\ reports 384 bytes, confirming both the layout and
Eq.~(\ref{eq:expansion}). The attacker selects the smallest currently-zero bit
whose flip covers the target, so the expansion is minimal.

\subsection{Real-runtime bypass and control}

Against the \FilC{} runtime, with the flip applied, the runtime permits the
previously out-of-bounds access in every scenario: an in-bounds access stays
ok, OOB read and OOB write both become \textbf{permitted}, and the runtime
object length reports $256\to384$\,B.  In the no-flip control, the same
victim binary and the same access path instead trigger a hard runtime panic
(``\texttt{cannot read/write pointer with ptr >= upper}'') and the victim is
killed before reporting, demonstrating that the access is genuinely
out-of-bounds and that only the metadata corruption removes the protection.

\subsection{Case studies}
\label{sec:casestudies}

We instantiate four victims, \texttt{generic}, \texttt{secret},
\texttt{authflag}, and \texttt{kv}, all compiled under \FilC{} and all defeated
by a single bit flip in the target object's upper-bound word. The four share one
skeleton, following the harness of \S\ref{sec:poc_real}: each is an ordinary,
\emph{correct} C program with no intentional memory error, relying entirely on
\InvisiCaps{} for spatial safety, that allocates the target, publishes its
runtime-assigned bounds, waits for the flip, and makes an access out-of-bounds for
the \emph{original} object. In each of the three cross-object cases the secondary
object is allocated from the same 256-byte size class and, because \FilC{}'s slab
allocator assigns consecutive increasing addresses, lands immediately above the
target (\S\ref{sec:phase2}), with the victim discarding any candidate that does
not, so the minimal upper-bound expansion reaches it.

\noindent\textbf{\texttt{generic}} isolates the raw primitive. It allocates one
256-byte object whose post-flip access reads and writes just past the original
end, both permitted under the corrupted bound and hard-panicked under the intact
one, establishing a forward, bounded out-of-bounds read/write window.
\textbf{\texttt{secret}} puts a confidential token in the neighbour, and the
expanded bound lets the target's own capability over-read across the allocation
boundary and disclose it, a direct information-disclosure primitive.
\textbf{\texttt{authflag}} is the write-side dual. Its neighbour is a non-admin
session record, and an out-of-bounds write sets that record's \texttt{is\_admin}
field from 0 to 1, after which the program reads the field through the
neighbour's own intact capability and elevates the user to admin.
\textbf{\texttt{kv}} is the realistic victim, a small multi-tenant credential
service built around real, unmodified SQLite compiled under \FilC{} and used
exactly as the project ships it for embedding (the single-file
\emph{amalgamation}). A guest request handler loads its own record into a heap
buffer while a second tenant's API key sits in an adjacent same-class ``vault''
buffer, and after a single-bit flip of the guest record's upper-bound word the
handler over-reads into the vault and discloses the admin's stored key. The
vulnerability is \emph{not} in SQLite: it arises entirely from the metadata
corruption in the service we built around it, showing the exposure reaches a
large, unmodified codebase, not just small purpose-built victims.

\subsection{Size-class generality}
\label{sec:sizeclass}

The case studies above use 256-byte target objects. To confirm the bypass is not
an artefact of a single allocator size class, we re-ran \texttt{generic},
\texttt{secret}, and \texttt{authflag} with 512-byte allocations, for which FUGC
assigns a different, exact-fit slot at a 528-byte stride. All three bypasses again
succeed against the runtime. Because the 512-byte class fits the request exactly
there is no allocator slack, so the secondary object's payload starts just 16\,B
past the original \texttt{upper} bound and less expansion is needed to reach it.
This shows up in the minimum flip bit, which drops to $k=6$ ($+64$\,B) for
\texttt{generic} and \texttt{authflag}, down from $k=7$ in the 256-byte case, and
is $k=7$ ($+128$\,B) for \texttt{secret}, tracking the class gap (stride minus
requested size) rather than the object's raw size. Every scenario still ends in
the same single-flip compromise as the 256-byte case.

\subsection{Physical feasibility of delivery}
\label{sec:phys_feas}

The case-study evaluation above delivers the flip out-of-band so that the
\emph{consequence} is measured independently of flip throughput.  To confirm
that the chain is nonetheless physically realisable,
we also exercised the bare-metal engine (\S\ref{sec:poc_phys}) on the Vendor~A
DDR3 lab machine, with no out-of-band assist anywhere.  Two outcomes follow,
with different reliabilities.

First, \emph{physical susceptibility}: the engine's self-templating pass hammers
attacker-owned memory with timing-selected same-bank double-sided aggressors and
reads back genuine, control-verified bit flips in both polarities, reproducing
reliably and confirming the module is susceptible before the victim is targeted.
Second, a \emph{full physical end-to-end}: via Flip-Feng-Shui page-frame
grooming~\cite{ffs} (\S\ref{sec:poc_phys}) the attacker donates known-flippy
frames, has the \FilC{} victim spray same-size-class objects until an
\InvisiCap{} upper-bound word lands on such a frame, and re-hammers the unchanged
aggressors.  When a real \zerotoone{} flip lands in a live victim's upper-bound
word, the runtime reports the object as larger and permits the out-of-bounds
read and write, observed directly through the same handshake channel and with no
simulation.  A successful campaign completes the bypass in roughly two minutes.

The two outcomes differ in reliability because the end-to-end path has a binding
stage the deterministic path lacks.  Landing a \emph{usable} flip in the exact
live word is probabilistic per run: the flip must (a)~be a \zerotoone{} flip,
(b)~on a bit currently zero in the stored word, (c)~increase rather than decrease
the bound, (d)~land in a low byte of the 8-byte word so the result stays a valid
pointer, and (e)~expand the bound far enough to reach the intended data.
Conditions (a)--(d) define an \emph{exploitable} window independent of the
target, and (e) is target-specific.  They compose, so a single grooming pass may
place no usable flip while a repeated campaign does, which is why the case-study
evaluation uses deterministic out-of-band delivery.

These observations are consistent with prior work establishing that single-bit
DDR3 flips are reproducible in tens of seconds~\cite{kim2014,seaborn2015} and
that TRR-equipped DDR4 falls to Blacksmith~\cite{blacksmith} and
RowPress~\cite{rowpress}.  The templating and grooming needed are the same
well-studied procedures used against page tables and other
targets~\cite{seaborn2015,saxena2023}.

\subsection{Physical feasibility under in-DRAM mitigation: DDR4}
\label{sec:ddr4_phys}

The DDR3 result above confirms the chain on a generation with no in-DRAM
defences.  DDR4 is a harder target, since every module of this class runs
in-DRAM TRR (\S\ref{section_background}). We therefore validate the complete
physical chain on a second, TRR-equipped lab machine, separate from the DDR3
systems, covering \emph{templating}, \emph{TRR-aware delivery}, and the
resulting \emph{consequence} in the \FilC{} runtime, with every stage physical
and no out-of-band step. The test machine is a commodity single-channel desktop
with 8\,GiB of memory, populated by one non-ECC DDR4 DIMM from Vendor~B, running
at the standard refresh rate with no way to disable in-package TRR. Consistent
with our disclosure policy (\S\ref{section_ethics_considerations}), we do not
name the vendor or platform model here, and will disclose the full configuration
once coordinated vendor disclosure permits.

\subsubsection*{Targeting under TRR}
We re-derive the same-bank conflict-timing oracle for this module
(\S\ref{sec:phase1}). Because the address mapping depends on the base address, the
engine assumes no fixed stride and instead empirically identifies the nearest
same-bank neighbours of each candidate victim cell, a step we confirmed necessary
because a stride transferred from the DDR3 module systematically misses the true
nearest row here. To evade TRR's activation sampler we use wide many-sided
hammering (\S\ref{sec:attack}), accompanying the true aggressor pair with a large
set of decoy rows that spread the sampler's limited tracking budget. Every
candidate flip is accepted only after a \emph{no-hammer control} on the identical
cell over the identical wait, rejecting any flip that also appears without
hammering as retention noise (the same discipline as \S\ref{sec:phys_feas}). The
cells on this module also exhibit \emph{variable retention time} (VRT),
susceptible only during transient weak-retention windows that open and close on a
timescale of minutes, so failure to reproduce a flip in the same cell later is
expected and does not invalidate an earlier control-verified result. Under this
discipline, self-templating in attacker-owned memory reliably yields
control-verified, hammer-induced \zerotoone{} and $1{\to}0$ flips, direct evidence
that the module's TRR implementation does not close the delivery channel.

\subsubsection*{Grooming under TRR: a zone-locality condition}
The end-to-end path uses the same Flip-Feng-Shui-style frame donation as
\S\ref{sec:phys_feas}, but a multi-zone allocator adds a constraint: a donated
frame is reused only if it lies in the zone the victim is currently allocating
from. We therefore have the victim perform a small ``scout'' allocation to reveal
the live zone, and confine templating and donation to that zone before the main
attempt. This is a general lesson for physical-frame grooming on modern multi-zone
Linux allocators, not a DDR4 peculiarity.

\subsubsection*{End-to-end result}
All four case studies (\S\ref{sec:casestudies}) were independently reproduced on
this module using a fully physical attack chain, with no out-of-band step. A
genuine hammer-induced flip was groomed onto the live \InvisiCap{} upper-bound
word, and the real \FilC{} runtime then permitted the corresponding
out-of-bounds access, observed directly through the handshake channel.
\texttt{generic} gained an out-of-bounds read and write, \texttt{secret}
disclosed the adjacent token, \texttt{authflag} escalated a non-admin session to
admin, and \texttt{kv}'s guest handler leaked the adjacent admin vault. Landing
the usable flip and completing the bypass took 10--35 minutes per run, the
spread reflecting the VRT-modulated, probabilistic character of the binding
stage.

As on DDR3, the binding step is landing a flip in the victim's exact live word,
subject to the same compound conditions as \S\ref{sec:phys_feas} and, here, the
zone-locality condition above, while templating and the \FilC{}-side consequence
remain reliable once the flip lands. This closes the objection that TRR-equipped
generations might shut the delivery channel even though the metadata exposure
remains: TRR as deployed on this module does not prevent a hammer-induced flip
from reaching and being honoured in a live victim's bounds metadata. We
demonstrate the bypass physically on both pre-TRR DDR3 and TRR-equipped DDR4, and
deterministically across all four case studies.

 \section{Discussion}
\label{sec:discussion}

\subsection{Why the attack works}

The exploit succeeds because of a fundamental abstraction mismatch. \FilC{}'s
correctness proof is stated at the C memory-model level, that no pointer
dereference can exceed its allocation, and holds within its layer, where every
dereference is instrumented and every bound checked. But it silently inherits the
physical layer's assumption that DRAM faithfully stores values. Rowhammer violates
this invisibly, with no system-call boundary crossed, no instruction touching the
target cell, and no signal or machine check raised. \FilC{} cannot detect it, so
the metadata-immutability assumption of \S\ref{sec:assumptions} is violated and the
safety proof collapses on a single physical cell.

\subsection{FUGC interaction}

\FilC{}'s concurrent collector FUGC is \emph{non-moving}
(\S\ref{section_background}), which works in the attacker's favour: once an
object is allocated and kept live, its \InvisiCap{} cell stays at the templated
physical location, with no compaction to contend with. The only collector-level
concern is \emph{reclamation}, since a freed slot may be reused and invalidate the
placement, so we retain live references to the groomed objects throughout and
observed no failures from relocation or slot reuse. A \emph{moving} collector
would reintroduce the relocation hazard and require pinning or renewed grooming.

\subsection{Limitations}

We are explicit about the scope of our evidence. The case-study evaluation models
delivery out-of-band for determinism, while the physical flip primitive is
established prior work that we separately confirmed end-to-end on two lab machines
(\S\ref{sec:phys_feas}, \S\ref{sec:ddr4_phys}), where the integrated path remains
probabilistic per run and on DDR4 is further constrained by TRR and allocator-zone
locality. These experiments establish the \emph{consequence}: once the metadata is
corrupted, the \FilC{} runtime is silently defeated regardless of how the flip was
delivered.

Our evaluation uses 256- and 512-byte victim objects (\S\ref{sec:sizeclass}) and
the inline-header layout of \FilC{}~0.679, and other size classes or future
versions may use different strides or header placements. The underlying exposure
remains structural, however: the bounds stored in DRAM are unauthenticated.
 \section{Related work}
\label{sec:related}

\subsection{Software and hardware memory safety.}
Retrofitting memory safety onto C/C++ spans compiler instrumentation and
hardware assistance.  SoftBound/CETS~\cite{softbound} provide complete spatial
and temporal safety by maintaining disjoint per-pointer metadata, and
AddressSanitizer~\cite{asan} catches many violations with red-zones and shadow
memory. Both keep their metadata in ordinary DRAM and are therefore, like
\FilC{}, exposed to a targeted hardware fault that alters that metadata without
tripping a check~\cite{kim2014}.

A natural comparison is CHERI~\cite{woodruff2014}, which presents two fault
targets. The first is the architecturally non-addressable one-bit validity tag on
each capability-sized location, which prevents software from forging capabilities.
An out-of-band \zerotoone{} fault in that tag storage could bypass this, but
whether Rowhammer can reach it is implementation-specific and we do not evaluate
it. The closer analogue to our attack is a flip in the addressable capability
encoding while the tag stays set. Here CHERI Concentrate uses a compressed,
floating-point-like encoding of base and top~\cite{CHERI_Concentrate}, so a single
flip does not simply add $2^k$ to the upper bound as in \FilC{}
(Eq.~\ref{eq:expansion}) but may narrow, widen, or malform the decoded bounds, and
the tag does not authenticate those bits. Whether one flip can yield a tagged,
well-formed, more-permissive capability must therefore be established
experimentally, so our \FilC{} result does not transfer to CHERI as a corollary
but motivates a concrete open question outside its usual reliable-hardware fault
model.

Arm MTE~\cite{mte} similarly assigns colours to allocations and checks them in
hardware while leaving allocation data in DRAM, and related primitives such as
Intel MPK, Arm's Permission Overlay Extension (POE/POE2), and virtualised MTE
control \emph{who} may access a region or \emph{how} it is coloured without
themselves authenticating DRAM-resident state against hardware faults. Any
unprotected bounds, tags, or permission metadata in commodity DRAM remains
exposed to an out-of-band flip.

\subsection{Rowhammer consequences}  A large body of prior work assumes a physical
bit flip can be induced and studies its \emph{consequences} for a particular
target. Prior flips have escalated kernel privileges via page-table
entries~\cite{seaborn2015}, corrupted RSA authentication state across co-located
VMs~\cite{ffs}, achieved deterministic exploitation on mobile devices through
predictable allocator behaviour~\cite{drammer}, and turned Rowhammer into a
\emph{read} primitive that leaks secrets without accessing them~\cite{rambleed}.
The work closest in spirit is \cite{anotherflip}, which flips instruction
\emph{opcodes} to turn one valid instruction into another and subvert security
checks. We apply the single-flip idea to bounds metadata, turning a valid
capability word into another valid, still-canonical word that encodes a wider
bound the runtime accepts. PT-Guard~\cite{saxena2023} protects page-table entries
with integrity MACs, and we recommend the same class of defence for
\InvisiCaps{}. Earlier compiler-based schemes likewise store bounds metadata in
DRAM and optimise it for performance~\cite{softbound,ZhengY:shadowboundehmptammacco}
without considering its integrity under hardware faults. Our contribution is
different in kind: a complete software memory-safety guarantee silently inherits
an unstated DRAM-integrity assumption, so a single flip in the very bounds it
enforces defeats the guarantee, and the defence's own metadata becomes the
bypass.

\subsection{Rowhammer techniques and defences.}
The flip primitive is robust and improving. Beyond the original DDR3
result~\cite{kim2014}, Rowhammer has been driven from
JavaScript~\cite{gruss2016,smash2021} and over the
network~\cite{throwhammer,nethammer}, and the in-DRAM TRR introduced with DDR4 has
been bypassed by TRRespass~\cite{trrespass}, Blacksmith~\cite{blacksmith},
RowPress~\cite{rowpress}, Half-Double~\cite{halfdouble}, and on AMD by
ZenHammer~\cite{zenhammer}. On the defence side, system-level SECDED ECC raises but
does not close the bar, since ECCploit~\cite{eccploit} and
Copy-on-Flip~\cite{didio2023} show that crafted multi-bit patterns still yield
silent corruption. DRAMA~\cite{pessl2016drama} provides the runtime bank-function
recovery our physical path uses to select genuine same-bank aggressors. We invoke
this work only to substantiate the antecedent of our conditional claim, that a
usable flip is attainable across DRAM generations, not to extend the Rowhammer
delivery literature itself.
 \section{Mitigations}
\label{sec:mit}

We group defences into two families. \emph{\FilC{}-specific} measures
authenticate or replicate in-DRAM bounds so that a silent flip becomes a
\emph{detected} fault, closing the exposure we demonstrate. \emph{General}
Rowhammer defences instead impede delivery and apply to any scheme that stores
metadata in DRAM. Preventing Rowhammer is a
hardware-level task, not \FilC{}'s burden: the specific measures make a flip
\emph{visible} to the runtime rather than stop it in silicon.

\subsubsection*{Cost and concurrency}
None of these defences is free. Authenticating or replicating bounds adds memory,
through wider headers or duplicate words, and cache pressure, on a runtime that
already incurs a several-fold overhead over native C~\cite{filc}. \emph{Per-use}
authentication prevents the silent use demonstrated here but charges every
dereference, whereas \emph{periodic} authentication, for example once per FUGC
sweep, amortises this but leaves the runtime honouring a corrupted bound until the
next check. Any ``verify-then-use'' check also has a time-of-check-to-time-of-use
(TOCTTOU) window unless verification and use are coupled tightly, and the design
must keep the protected word, any replica or authenticator, and any key consistent
under concurrent FUGC, which traverses and rewrites free-list state. These are
engineering efforts, not drop-in patches.

\subsection{\FilC{}-specific: authenticate the metadata}

\subsubsection*{Keyed capability MACs} At allocation time, compute and store
$\mathrm{MAC}_K(\texttt{lower}\ \Vert\ \texttt{upper})$ for each
object's bounds, and verify it before use. A flipped upper-bound word
no longer matches the stored tag, so the dereference faults rather than silently
accepting the expanded bound. PT-Guard~\cite{saxena2023} applies the same defence
to page tables, and PACMem~\cite{LiY:pacmemesatmsvapa} and
CryptSan~\cite{HohentannerK:crypsanlapafmsicc} repurpose Arm Pointer Authentication
to protect pointers and metadata. If Arm's Pointer Authentication Codes are the
authenticator, they are speculatively forgeable~\cite{PACMAN} and a MAC over the
bounds inherits that issue. Permission-overlay primitives such as Arm POE/POE2
might support related protection. Three design constraints are easy to overlook.

\begin{itemize}

\item \emph{No free ``spare word.''} \FilC{}~0.679's 16-byte header is fully
  occupied by the 8-byte \texttt{upper} and \texttt{aux} words
  (\S\ref{sec:layout}). Even a truncated tag widens it, perturbing the
  \texttt{libpas} size classes and slab stride (\S\ref{sec:phase2}) and changing
  ABI-visible layout. This is a genuine layout change, not reuse of an idle field.

\item \emph{The key must live off the heap.} If $K$ sits in ordinary DRAM it
  shares the metadata's fault exposure, and an attacker who recovers or forces
  $K$ could forge tags. A sound design keeps $K$ in a root secret held in a
  model-specific or control register outside hammerable DRAM, since a reserved
  general-purpose register is visible to other code and may be spilled.

\item \emph{Verification is not ``a few instructions.''} A software
  SipHash~\cite{siphash} over the bound pair costs tens of cycles per check on a
  stock ISA. Amortising it across FUGC sweeps improves throughput but detects
  corruption only at the next collection.

\end{itemize}

\subsubsection*{Redundant bound storage} Store \texttt{upper} at two physically
distant locations and compare them at each use, forcing two independent flips of
identical effect and making exploitation substantially harder. The cost is an
extra cache miss per dereference and a doubled bound footprint, and the replicas
must stay atomically consistent across legitimate updates and FUGC or the
comparison itself may fault.

\subsubsection*{A runnable check}
We implemented a self-contained program that reproduces the
\FilC{}~0.679 inline-header layout (\S\ref{sec:layout}) and implements three
variants of the bounds check: the current unauthenticated baseline, the keyed
capability MAC, and redundant bound storage. Given the identical
single-bit flip applied to the deterministic path, the \FilC{}
baseline silently accepts the corrupted bound. In contrast, both the
keyed MAC and the redundant copy trigger a fault.  This isolates
metadata integrity from physical delivery.

\subsection{General Rowhammer defences: raise the delivery bar}

These measures are not specific to \FilC{}: they protect any in-DRAM-metadata
scheme, but \FilC{} cannot deploy them on its own.

\subsubsection*{ECC memory} Both lab machines use non-ECC memory, and on DDR4
in-DRAM TRR did not prevent delivery (\S\ref{sec:ddr4_phys}), so it cannot
substitute for ECC. SECDED ECC corrects an isolated single-bit error and would
prevent software from observing the corruption used here, but does not
categorically prevent crafted \emph{multi-bit} patterns. ECCploit~\cite{eccploit}
turns correction latency into a timing side channel for finding exploitable
multi-bit templates, and defeating system-level ECC requires additional ECC-aware
templating and fault handling~\cite{didio2023}. ECC thus remains a
\emph{delivery}-side defence that makes corruption harder to deliver but does not
protect the runtime once corrupted metadata is consumed, which motivates the
metadata-authentication defences above.

\textbf{Physical row isolation.} Keeping attacker and victim pages in
non-adjacent DRAM rows prevents disturbance from reaching the metadata and
therefore denies Phase~3. Software mitigations include ZebRAM, which interleaves
guard rows~\cite{KonothRK:zebramcacspara}, and SoftTRR, which uses a software
analogue of TRR to protect page tables~\cite{ZhangZ:softTRRpptarausotrr}.
Neither is widely deployed on commodity Linux, and both add memory and
performance overheads.

\textbf{clflush restriction.} Blocking \texttt{clflush} at the hypervisor
raises the cost of hammering by forcing slower software eviction sets, but does
not eliminate it: JavaScript-based attacks hammer without
\texttt{clflush}~\cite{gruss2016,smash2021}.

 \section{Conclusion}

We showed that \FilC{}'s \InvisiCap{} memory safety rests on an unstated
assumption, that commodity DRAM faithfully preserves the bounds metadata the
runtime checks. In \FilC{} 0.679 that metadata is an unauthenticated word of
ordinary DRAM at \payloadhdr{}, and a single \zerotoone{} flip in it, the
canonical effect of a Rowhammer fault, silently expands the object's permitted
range by $2^k$ bytes while remaining invisible to the runtime. We demonstrated the
consequence end-to-end against the \FilC{}~0.679 runtime, the no-flip control
hard-panicking as expected, with a single flip triggering each of our four case
studies, from an out-of-bounds read/write primitive to cross-tenant key disclosure
from a multi-tenant SQLite service. We also validated the complete physical chain,
landing a genuine hammer-induced flip in a live victim's metadata word on a
susceptible DDR3 machine and reproducing it for all four case studies on a
TRR-equipped DDR4 module, so in-DRAM Target Row Refresh did not close the delivery
channel on the tested module. Our results are best understood as a conditional
claim:
\emph{if templating and memory placement deliver a usable Rowhammer flip to the
stored upper-bound word of a live object, then \FilC{} can silently permit an
out-of-bounds access to that object, without a panic, signal, or machine check,
thereby defeating \FilC{}'s spatial-safety guarantee for the object.} The
Rowhammer literature has already established the feasibility of its antecedent
across DRAM generations.

\section{Ethics Considerations}\label{section_ethics_considerations}

We have considered the ethical implications of this work following widely
accepted ethical principles for security research, in particular the Menlo
Report principles of beneficence, respect for persons, and justice.

\noindent\textbf{No harm, benign by design.} Every experiment runs entirely on
our own lab machines. The attacker process forks the victim as its own child and
communicates only through a private shared-memory handshake, so no other process,
user, host, or remote system is targeted and no human-subject, personal, or
production data is involved. The bare-metal engine disturbs only DRAM we own, on
dedicated, isolated machines with no network exposure. The proof-of-concept code
grants no capability its operator does not already possess, has no self-propagation
or network component, and cannot target a process the operator does not own. The
deterministic path writes only to a forked child the operator already controls (via
\texttt{/proc/\textit{pid}/mem}, permitted for one's own child under the default
Linux \texttt{Yama} policy). We withhold the exact platform and DRAM module
identities (\S\ref{sec:ddr4_phys}) pending coordinated vendor disclosure.

\noindent\textbf{Nature of the weakness.} What we report is not an implementation
defect that a point patch removes, but a structural consequence of storing
unauthenticated bounds metadata in commodity DRAM, a deliberate \FilC{} design
choice for ISA compatibility and complete checking on stock CPUs. The delivery
primitive, Rowhammer, is well-established prior work and we contribute no new or
faster way to induce flips, so the marginal risk of publication is low while the
benefit, alerting designers of compiler-based memory-safety systems to an
unguarded assumption, is substantial.

\noindent\textbf{Responsible disclosure.} Ahead of publication we notify, on one
coordinated timeline, the \FilC{} maintainer (the exposure is specific to
\InvisiCap{}'s design, and we share our proposed mitigations, \S\ref{sec:mit}) and
the DRAM manufacturers of the modules on which we demonstrate a physical flip. We
contacted the maintainer and one affected manufacturer on 27 July 2026 without
response. On 12
August 2026 we filed a report with the CERT Coordination Center (CERT/CC) via its
VINCE portal and shared an anonymised draft under embargo. On 14 August 2026
CERT/CC declined to open a case or pursue a CVE, reasoning that Rowhammer is
already public and the exposure is not a code defect. Following its recommendation
we are setting a public-disclosure date, sending
dated advisories to the affected parties, and will request a CVE from MITRE if
they remain unresponsive. Given that the exposure follows a documented design
decision rather than a hidden bug, we judged a coordinated-disclosure window
rather than a strict embargo appropriate.

\bibliographystyle{IEEEtran}
\bibliography{references}

\end{document}